\documentclass[12pt]{article}
\usepackage{lnfprep}
\usepackage{epsfig}

\def\c2{CLEO~II.V}

\def\d0d0{ D^0\bar{D}^0 }
\def\p0p0{ P^0\bar{P}^0 }
\def\qp2{ \Bigl| \frac{q}{p} \Bigr|^2 }
\def\pq2{ \Bigl| \frac{p}{q} \Bigr|^2 }

\def\be{\begin{equation}}
\def\ee{\end{equation}}
\def\bea{\begin{eqnarray}}
\def\eea{\end{eqnarray}}

\newcommand{\mevcc}{\ensuremath{\rm{MeV}/c^2}\,}

\newcommand{\gevc}{\ensuremath{\mathrm{GeV}/c}\,}

\newcommand{\AmS}{{\protect\the\textfont2
  A\kern-.1667em\lower.5ex\hbox{M}\kern-.125emS}}

\newcommand{\Header}{
  \begin{tabular}{rl}
  \hspace{-.4cm}\includegraphics{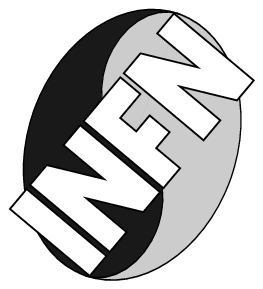} &
    \renewcommand{\arraystretch}{0.5}
    \begin{tabular}{r}
      {\hspace{1cm}~\LARGE\sffamily LABORATORI~ NAZIONALI~ DI~ FRASCATI}\\
      \\
      {\Large\sffamily SIS-Pubblicazioni}\\
    \end{tabular}
    \renewcommand{\arraystretch}{1}
  \end{tabular}
  \vskip 1cm
  \begin{flushright}
  \renewcommand{\arraystretch}{0.5}
    \begin{tabular}{r}
      {\underline{LNF-02/029 (P)}}\\    
      {\small 10 Dicembre 2002} \\      
      \\
    \end{tabular}
  \end{flushright}
  \renewcommand{\arraystretch}{1}
  \vskip 0.5 cm
  }

\begin{document}
\begin{titlepage}
\title{
  \Header
  {\large \bf Light Quark Spectroscopy Results from FOCUS and E687}
}
\author{ Stefano Bianco \\
 Laboratori Nazionali di Frascati  
via E.~Fermi 40, Frascati 00044, Italy      \\
         on behalf of the FOCUS and E687 Collaborations
         \thanks{\scriptsize Co-authors are 
 P.L.~Frabetti ({\bf INFN and Bologna}), 
J.M. Link,
V.S. Paolone, 
M. Reyes, 
P.M. Yager 
 ({\bf UC DAVIS}); 
J.C. Anjos,
I. Bediaga, 
C. G\"obel, 
J. Magnin, 
A. Massafferri,
J.M. de Miranda, 
I.M. Pepe, 
A.C. dos Reis,
F. Sim\~ao 
 ({\bf CPBF, Rio de Janeiro});
S. Carrillo, 
E. Casimiro, 
H. Mendez, 
\hbox{A.S\'anchez-Hern\'andez,},
C. Uribe, 
F. Vasquez 
 ({\bf CINVESTAV, M\'exico City});
L. Agostino,
L. Cinquini, 
J.P. Cumalat, 
C. Dallapiccola,
J.F. Ginkel,
J.E. Ramirez, 
B. O'Reilly, 
I. Segoni,
E.W. Vaandering 
 ({\bf CU Boulder});
J.N. Butler, 
H.W.K. Cheung,
G. Chiodini, 
S. Cihangir,
I. Gaines, 
P.H. Garbincius, 
L.A. Garren,
E. Gottschalk,     
S.A. Gourlay, 
D.J. Harding,
P.H. Kasper,
A.E. Kreymer, 
R. Kutschke,
P. Lebrun,
S. Shukla,
M. Vittone
 ({\bf Fermilab}); 
R. Baldini-Ferroli,
L. Benussi, 
F.L. Fabbri, 
A. Zallo 
 ({\bf INFN Frascati}); 
C. Cawlfield, 
R. Culbertson,
R. Greene,
D.Y. Kim,
K.S. Park, 
A. Rahimi,
J. Wiss 
 ({\bf UI Champaign}); 
R. Gardner,
A. Kryemadhi
 ({\bf Indiana }); 
Y.S. Chung,
J.S. Kang, 
B.R. Ko, 
J.W. Kwak,
K.B. Lee, 
S.S. Myung
 ({\bf Korea University, Seoul}); 
 K.Cho, H.Park ({\bf Kyungpook National University, Taegu});
G. Alimonti,
S. Barberis, 
A. Cerutti,
G. Bellini,
M. Boschini, 
D. Brambilla,
B. Caccianiga, 
A. Calandrino, 
L. Cinquini,
P. D'Angelo, 
M. DiCorato, 
P. Dini, 
L. Edera, 
S. Erba,
M. Giammarchi,
P. Inzani,
F. Leveraro, 
S. Malvezzi, 
D. Menasce, 
E. Meroni,
M. Mezzadri, 
L. Milazzo, 
L. Moroni,
D. Pedrini,     
L. Perasso,
C. Pontoglio,
F. Prelz, 
M. Rovere, 
A. Sala,
S. Sala 
D. Torretta
 ({\bf INFN and Milano}); 
D. Buchholz,
D. Claes,
B. Gobbi 
 ({\bf Nortwestern});
J.M. Bishop,
N.M. Cason,
C.J. Kennedy,
G.N. Kim,
T.F. Lin,
D.L. Puseljic,
R.C. Ruchti,
W.D. Shephard,
J.A. Swiatek,
Z.Y. Wu 
 ({\bf Notre Dame});
T.F. Davenport III 
 ({\bf UNC Asheville});
V. Arena,
G. Boca, 
G. Bonomi, 
C. Castoldi,
G. Gianini, 
G. Liguori, 
M. Merlo, 
D. Pantea,
S.P. Ratti, 
C. Riccardi,
P. Torre, 
L. Viola, 
P. Vitulo 
 ({\bf INFN and Pavia});
H. Hernandez, 
A.M. Lopez, 
L. Mendez,
A. Mirles, 
E. Montiel, 
D. Olaya, 
J. Quinones, 
C. Rivera, 
Y. Zhang 
 ({\bf Mayaguez, Puerto Rico});
N. Copty, 
M. Purohit, 
J.R. Wilson 
 ({\bf USC Columbia});
K. Cho, 
T. Handler 
 ({\bf UT Knoxville}); 
D. Engh, 
W.E. Johns, 
M. Hosack,
M.S. Nehring, 
M. Sales, 
P.D. Sheldon,
K. Stenson, 
M.S. Webster 
 ({\bf Vanderbilt}); 
M. Sheaff 
 ({\bf Wisconsin, Madison}); 
Y. Kwon 
 ({\bf Yonsei University, Korea}).
         }
        }
\maketitle
\baselineskip=14pt

\begin{abstract}
 Using the FOCUS spectrometer (experiments 687 and 831 at Fermilab) 
 we confirm the existence of a diffractively photoproduced
 enhancement in 
 $K^+K^-  $ at 1750~\mevcc with nearly 100 times the statistics of previous
 experiments. 
 We also observe a narrow dip structure at 1.9~GeV/c$^2$
 in a study of diffractive photoproduction of
 the $~3\pi^+3\pi^-$ final state.
\end{abstract}
\vfill
\begin{flushleft}
  \vskip 0.1cm
{PACS: 14.65.Bt; 13.60.-r; 12.39.Mk   }  
\end{flushleft}
\begin{center}
Presented at the Int. Conf. on High Energy Physics (ICHEP02), Amsterdam, The
Netherlands, August 2002.
\end{center}
\vfill
\end{titlepage}
\pagestyle{plain}
\setcounter{page}2
\baselineskip=17pt

In this paper we present results on the spectroscopy of high-energy light
meson photoproduction from the FOCUS (E831) and E687 experiments at
Fermilab. Although both experiments were focused on charm physics, a very
large sample of diffractively photoproduced light-meson events was also
recorded. Details on the detector can be found in
Ref.~\cite{Link:2002mp}, Ref.~\cite{Frabetti:2001ah},
and references therein.  
\section{Observation of a 1750~\mevcc Structure in the Diffractive
Photoproduction of $K^+K^-$} 
The data for this analysis was collected by the Wideband photoproduction
experiment FOCUS during the Fermilab 1996--1997 fixed-target run.  In
FOCUS, a forward multi-particle spectrometer is used to 
measure the interactions of high energy photons on a segmented BeO target. 
A photon beam is
derived from the bremsstrahlung of secondary electrons and positrons
with an $\approx 300$ GeV endpoint energy produced from the 800
GeV/$c$ Tevatron proton beam.
The FOCUS detector is a large aperture, fixed-target spectrometer with
excellent vertexing, particle identification, and
reconstruction capabilities for photons, $\pi^0$'s, and $K_S$.

\begin{figure}
 \centering
  \includegraphics[width=6.0cm,height=4.0cm]{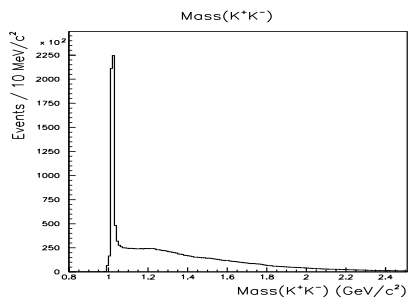}
 \caption{The $K^+K^-$ sample with no cut on $p_T$.}
 \label{fig_1}
\end{figure}

\begin{figure}
 \centering
  \includegraphics[width=7.0cm,height=7.0cm]{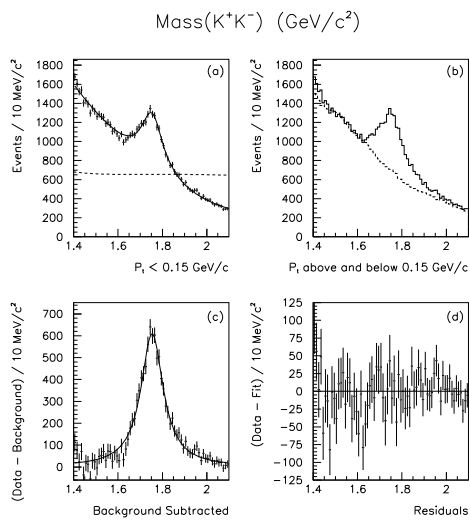}
 \caption{(a) The $K^+K^-$ mass spectrum with the requirement that $p_T <
{\rm 0.15}$~\gevc.  The spectrum is fit with a non-relativistic
Breit-Wigner distribution and a quadratic background.  The dotted line is
the Monte Carlo efficiency on a scale from 0 to 100${\rm \%}$.  (b) The
solid line is the $K^+K^-$ mass spectrum with the requirement that $p_T <
{\rm 0.15}$~\gevc.  The dotted line is the $K^+K^-$ mass spectrum with $p_T
> {\rm 0.15}$~\gevc scaled to the size of the low $p_T$ spectrum for
comparison.  (c) The data and fit after subtracting the quadratic
polynomial background shape.  (d) The data minus the fit.} 
 \label{fig_3}
\end{figure}

Our sample of $K^+K^-$ events, selected using the criteria described in
detail in Ref.~\cite{Link:2002mp}, 
shows a clear $\phi (1020)$ signal dominating the spectrum
(Fig.~\ref{fig_1}).  The 
diffractive component of the production of the $\phi (1020)$ shows up as a
peak in the $p_T$ spectrum.
Cutting around this peak by
requiring $p_T < {\rm 0.15}$~\gevc, we select a diffractive sample of
$K^+K^-$ events, in which a clear enhancement appears in the mass spectrum
near 1750~\mevcc (Fig.~\ref{fig_3}.a).  Figure ~\ref{fig_3}.b  confirms
that the enhancement 
appears at only low $p_T$.  Plotting the $p_T$ spectra in the 1750 region
(1640--1860~\mevcc) and in the two sideband regions (1500--1600~\mevcc and
1900--2100~\mevcc), it is seen that the 1750 region has a peak in the $p_T$
spectrum in nearly the same place as the $\phi(1020)$ peak, but the
sideband regions have significantly smaller $p_T$ peaks,
indicating that the background under the X(1750) signal is largely
non-diffractive. 
Fitting the 1750~\mevcc mass region with a non-relativistic Breit-Wigner
distribution and a quadratic background, we find 
\begin{displaymath}
\textrm{Yield} = 11,700\pm 480~{\rm Events}
\end{displaymath}
\begin{displaymath}
\mathrm{M} = 1753.5\pm 1.5\pm 2.3~\mevcc
\end{displaymath}
\begin{displaymath}
\mathrm{\Gamma} = 122.2\pm 6.2\pm 8.0~\mevcc
\end{displaymath}
%
%
There is a region near 1600~\mevcc where there is some discrepancy in our
fit to the $K^+K^-$ mass spectrum.  The residuals show that the statistical
significance of this discrepancy is not strong (Fig.~\ref{fig_3}.d).   
Finally, we searched for the X(1750) enhancement in K*K and we find
\begin{eqnarray*}
 \frac{\Gamma (X(1750) \rightarrow  \overline{K}^{*0}K^0 \rightarrow
 K^{-}\pi^{+}K_S+c.c.)}{\Gamma (X(1750) \rightarrow K^+K^-)}   & & \\
 <0.065~{\rm at}~90 \% ~{\rm C.L.}  & &
\end{eqnarray*}
\begin{eqnarray*}
 \frac{\Gamma (X(1750) \rightarrow  K^{*+}K^{-} \rightarrow
 K_S\pi^{+}K^{-}+c.c.)}{\Gamma (X(1750) \rightarrow K^+K^-)} & & \\
<0.183~{\rm at}~90 \% ~{\rm C.L.}   & &
\end{eqnarray*}
The two relative branching ratios were measured
to be $-0.083 \pm 0.081$ and $0.065 \pm 0.072$, respectively. 
Because of the large discrepancies in mass and relative branching fractions
to $K^+K^-$ and $K^*K$, we do not believe it is reasonable to identify the
X(1750) (which should be assigned the photon quantum numbers J$^{\rm
PC}=1^{--}$) with the $\phi (1680)$.  In fact, because the mass of the
X(1750) is significantly higher than all known vector mesons, the most
massive of which are the $\omega (1650)$, $\phi (1680)$, and $\rho (1700)$, an
interpretation claiming the X(1750) is some combination of interfering
vector mesons also seems highly unlikely.  The interpretation of the
X(1750) remains uncertain. 
\section{A Narrow Dip in A Study of Diffractive Photoproduction}
\begin{figure}
 \centering
  \includegraphics[height=3.5cm,width=6.0cm]{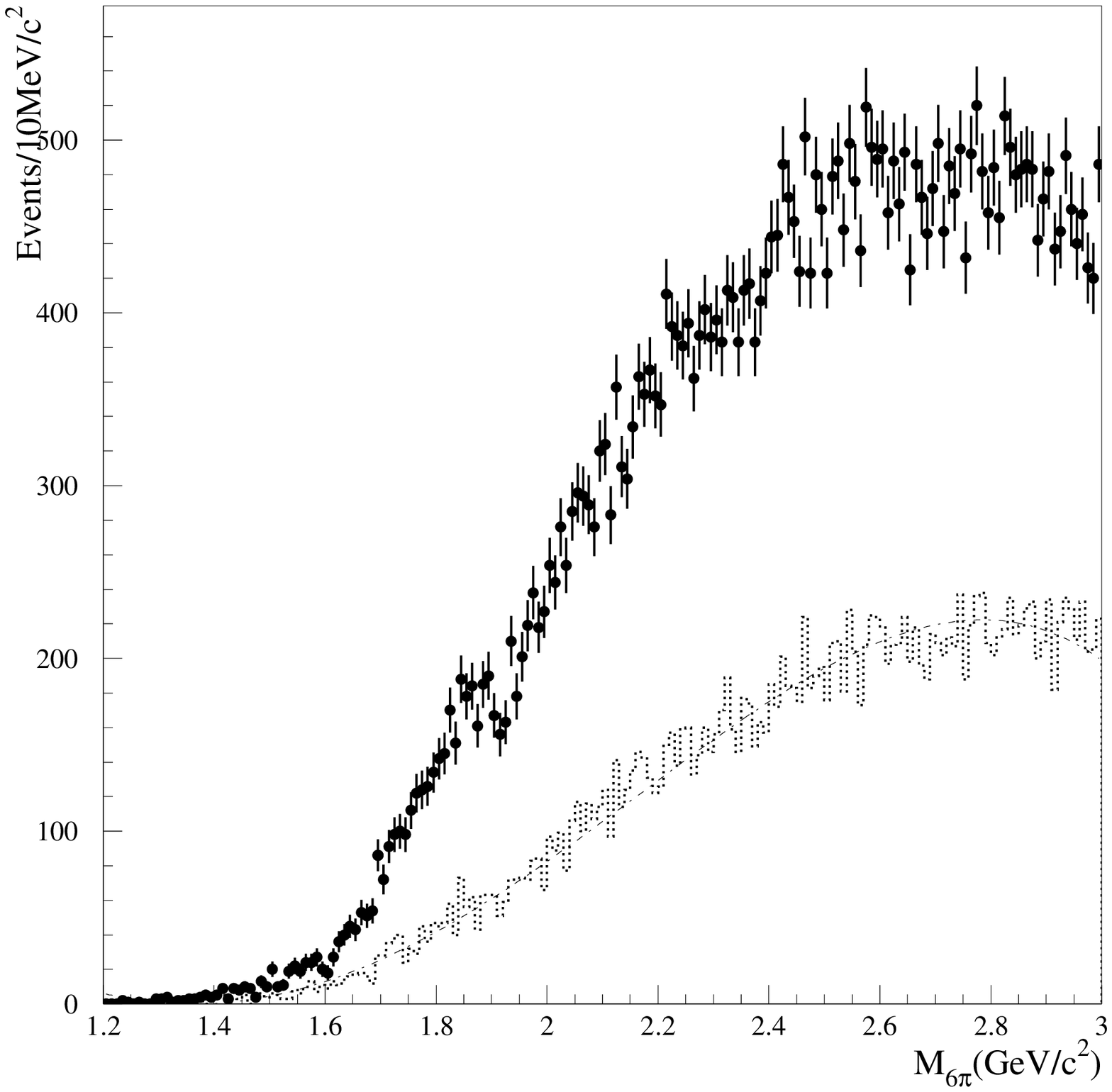}
  \includegraphics[height=3.5cm,width=6.0cm]{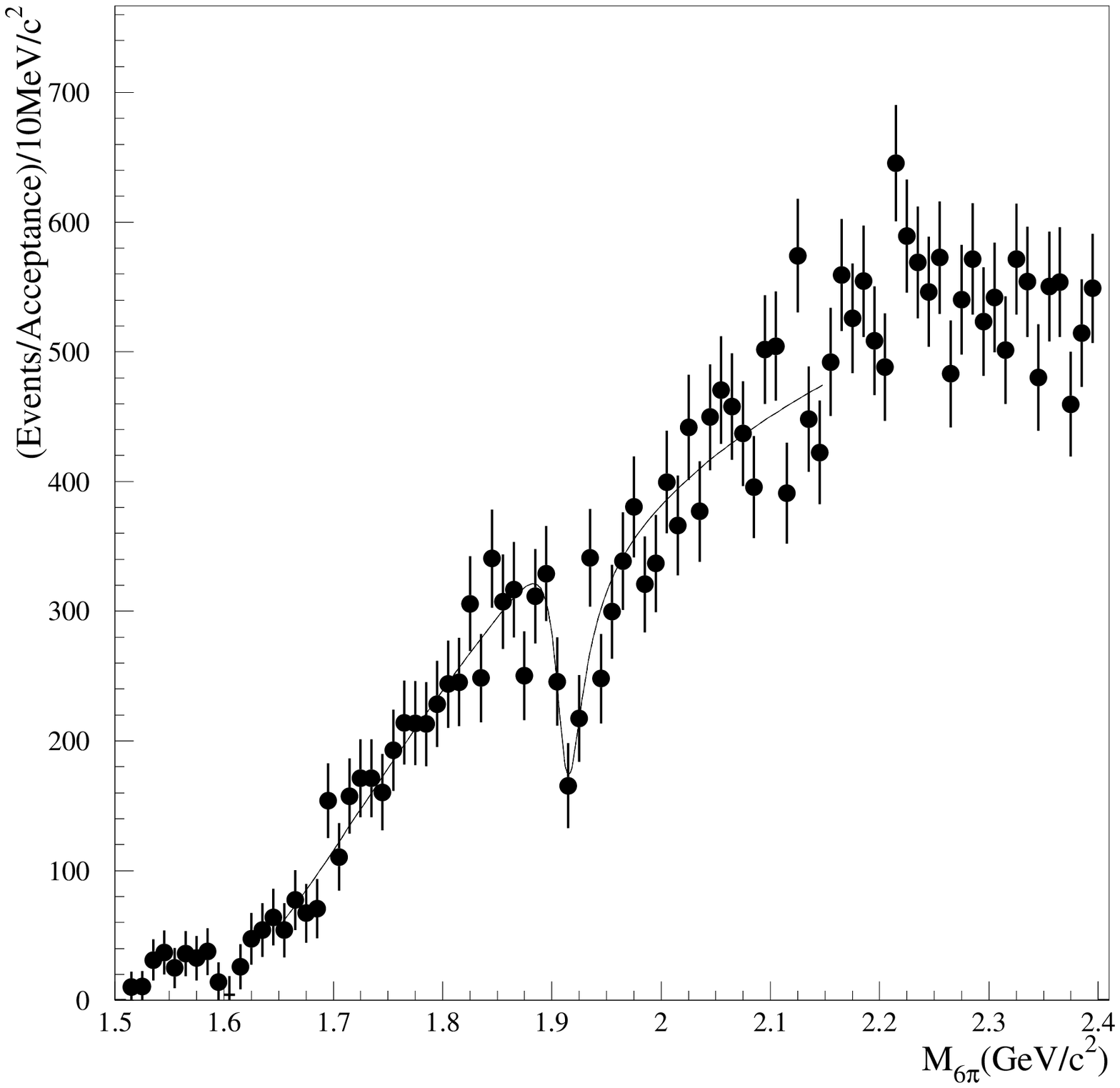}
\caption{ a) Distribution of $3\pi^+ 3\pi^- $ invariant mass in the
1.2$-$3.0~ GeV/c$^2$ mass
range: coherent plus incoherent contribution.
Dotted distribution: incoherent contribution. b) Acceptance-corrected
distribution of $3\pi^+ 3\pi^- $ invariant
mass for diffractive events.
The mass resolution has been unfolded. }
\label{fig:third}
\vfill
\end{figure}

The Fermilab experiment 687, FOCUS predecessor,  collected
data during the
1990/91 fixed-target runs at the Wideband Photon beamline at
Fermilab. 
We report on a study
of the diffractive photoproduction of the $3\pi^{+}3\pi^{-}$ final state
and the
observation of a narrow dip in the mass spectrum at 1.9~GeV/c$^2$.
Pions are produced in photon interactions in the Be target.
The data acquisition trigger requires a minimum energy deposition in the
hadron
calorimeters
located behind the electromagnetic calorimeters and
at least three charged tracks outside the pair region.
Details of event selection, analysis cuts and fitting strategies are
reported in \cite{Frabetti:2001ah}. The invariant mass distribution of
diffractively produced  $3\pi^{+}3\pi^{-}$ final states shows a dip
structure at about 1.9~GeV/c$^2$. No evidence for structures is shown in
the incoherent $(P_T^2>0.040 {\rm GeV/c}^2$ subsample (Fig.\ref{fig:third}a).
Fig.~\ref{fig:third}b) shows the coherent mass distribution after
correcting for efficiency and acceptance, and unfolding the spectrometer
mass resolution. The dip structure has been characterized by a
two-component fit, adding coherently a relativistic Breit-Wigner resonance
to a diffractive continuum contribution. Fit values show consistent
evidence for a narrow resonance at M$_r$ ~=~1.911 $\pm$ 0.004 $\pm$ 0.001
~GeV/c$^2$ with a width
$\Gamma$ = 29 $\pm$ 11 $\pm$ 4 ~MeV/c$^2$,
where the errors quoted are statistic and systematic,
respectively. Such a resonance could be
assigned the photon quantum numbers (J$^{\rm PC}=1^{--}$) and G=+1, I=1 due
to the final state multiplicity.
There is little
understanding of the specific mechanism responsible for this destructive
interference.
\par
\vfill
\section{Conclusions}
 We have reported on new results from fixed target photoproduction at
 Fermilab, which show how a high-resolution multiparticle spectrometer,
 coupled to a high-energy photon beam, constitutes a tremendously effective
 tool to explore the light vector meson spectroscopy and dynamics. More
 results will be coming soon.

\vfill

\end{document}